\documentclass[useAMS,usenatbib]{mn2e}
\newcommand{\teff}{\mbox{$T_{\rm eff}$}}
\newcommand{\logg}{\mbox{$\log g$}}
\newcommand{\halpha}{\mbox{$H_\alpha$}}
\newcommand{\hbeta}{\mbox{$H_\beta$}}
\newcommand{\vsini}{\mbox{$v \sin i$}}
\newcommand{\mactrb}{\mbox{$v_{\rm mac}$}}
\newcommand{\mictrb}{\mbox{$v_{\rm mic}$}}
\newcommand{\loggf}{\mbox{$\log gf$}}
\newcommand{\kms}{\mbox{km\,s$^{-1}$}}
\usepackage[dvips]{graphicx}

\title[Parameters of WASP planet host stars]
 {Accurate spectroscopic parameters of WASP planet host stars\thanks{The analysis is based on data obtained through observing programs 072.C-0488, 082.C-0040, 082.C-0608,
283.C-5017, 084.C-0185 \& 087.C-0649, from ESO's HARPS spectrograph, mounted at the 3.6 m at La Silla, Chile.}}
\author[A.P.~Doyle et al.] {Amanda P.~Doyle$^{1}$\thanks{E-mail: a.doyle@keele.ac.uk},  B.~Smalley$^{1}$, P.F.L.~Maxted$^{1}$, D.R.~Anderson$^{1}$, \newauthor A.~Collier~Cameron$^{2}$, M.~Gillon$^{3}$, C.~Hellier$^{1}$, D.~Pollacco$^{4}$, D.~Queloz$^{5}$, \newauthor A.H.M.J.~Triaud$^{5}$, R.G. West$^{6}$\\
$^{1}$ Astrophysics Group, Keele University, Staffordshire ST5 5BG, UK\\
$^{2}$ SUPA, School of Physics and Astronomy, University of St. Andrews, North Haugh, Fife, KY16 9SS, UK\\
$^{3}$ Universit\'e de Li\`ege, Allée du 6 ao\^ut 17, Sart Tilman, Li\`ege 1, Belgium\\
$^{4}$ Astrophysics Research Centre, School of Mathematics \& Physics, Queen's University, University Road, Belfast, BT7 1NN, UK\\
$^{5}$ Observatoire de Gen\`eve, Universit\'e de Gen\`eve, 51 chemin des Maillettes, 1290 Sauverny, Switzerland\\
$^{6}$ Department of Physics and Astronomy, University of Leicester, Leicester, LE1 7RH, UK}

\date{Released 2012 Xxxxx XX}

\pagerange{\pageref{firstpage}--\pageref{lastpage}} \pubyear{2012}

\def\LaTeX{L\kern-.36em\raise.3ex\hbox{a}\kern-.15em
    T\kern-.1667em\lower.7ex\hbox{E}\kern-.125emX}

\begin{document}

\label{firstpage}

\maketitle

\begin{abstract}
We have made a detailed spectral analysis of eleven Wide Angle Search for Planets (WASP) planet host stars using high signal-to-noise (S/N) HARPS spectra. Our line list was carefully selected from the spectra of the Sun and Procyon, and we made a critical evaluation of the atomic data. The spectral lines were measured using equivalent widths. The procedures were tested on the Sun and Procyon prior to be being used on the WASP stars. The effective temperature (\teff), surface gravity (\logg), microturbulent velocity (\mictrb) and metallicity were determined for all the stars. We show that abundances derived from high S/N spectra are likely to be higher than those obtained from low S/N spectra, as noise can cause the equivalent width to be underestimated. We also show that there is a limit to the accuracy of stellar parameters that can be achieved, despite using high S/N spectra, and the average uncertainty in \teff, \logg, \mictrb\ and metallicity is 83 K, 0.11 dex, 0.11 \kms\ and 0.10 dex respectively.

\end{abstract}


\section{Introduction}
Since the first exoplanet was discovered orbiting 51 Pegasi \citep{mq95}, hundreds more have been detected around other stars. Due to biases inherent in the detection techniques, the majority of the planets discovered to date are giant planets with short orbital periods. However, with the advent of the \emph{Kepler} space mission \citep{bo11}, planets with properties more akin to Earth are now being discovered. Accurate planetary parameters are essential, whether one is trying to study hot Jupiters or find another planet like Earth. In the case of transiting planets, information such as the orbital period can be extracted directly from the light curve, however the mass and radius of the planet are coupled with the mass and radius of the host star. The transit of an exoplanet across its host star will only yield the ratio of the planet to star radius and the mass of a planet, acquired from combining transiting and radial velocity data, cannot be determined independently from the mass of the star \citep{w10}.

It is thus imperative that the mass and radius of the host star are known precisely, but obtaining direct measurements is only possible for a limited number of stars. For instance, a fundamental value of mass cannot be obtained unless the star is in a binary system \citep*{twh08}, and determining the radius of a star requires knowledge of the angular diameter, which in turn needs a known distance \citep{n07}. Most current distance measurements were acquired using the parallax technique with ESA's \emph{Hipparcos} satellite, and these will be complemented in the near future by data from ESA's \emph{Gaia} mission, due to launch in 2013 \citep*{gss10}. 

In the absence of direct measurements, stellar spectroscopy can be utilised to determine the effective temperature (\teff), the surface gravity (\logg), and metallicity, [Fe/H]\footnote{Abundances are given in the format of $\log(A)$ + 12, where $\log(A)$ is the number ratio of the element with respect to hydrogen, log~($N_{el}$/$N_{H}$). When square brackets are used, e.g. [Fe/H], it denotes the abundance relative to the solar value, where as $\log A$(El) indicates $\log(A)$ + 12. Using curved brackets, (Fe/H), can also denote $\log (A)$ + 12}, of the planet host star. These parameters are then used to infer the stellar mass and radius, based on a calibrations such as \citet*{tag10}, or a grid of stellar models such as \citet{gi00}.

Knowledge of the chemical composition of the host stars has additional value, as planets are more likely to be found around high metallicity stars (\citealt{go98}; \citealt*{sim04}; \citealt{fv05}). It has also been suggested that stars with planets can have different chemical compositions to stars without known planets \citep{m09}.

In order to extract the maximum amount of information from the statistics of planet frequency and stellar abundances, these abundances need to have been measured with the highest level of precision possible. However, determining abundance is by no means an exact process. They are heavily influenced by the reliability of atomic data and the quality of the spectra. 

Details about the electron transitions that cause spectral lines, along with data on how these lines can be broadened by atomic processes, ideally need to be obtained via laboratory measurements. However, the sheer multitude of spectral lines makes this a lengthy and difficult task, so that often atomic data is inferred from indirect methods. The result of this is that even spectra with a high signal-to-noise ratio (S/N) there is a limit as to how precisely one can measure elemental abundances in stars. 

Spectral analysis of low S/N spectra is no easy task, as the excessive noise in the continuum of these spectra makes normalisation difficult, causes the wings of strong lines to be underestimated, and depletes the number of weak lines available for measurement. The initial analyses for the WASP planet host stars for the discovery papers are usually performed using coadded spectra from the CORALIE spectrograph, which has a resolution between 55\,000 and 60\,000 (\citealt{q01}; \citealt{wi08}). These spectra typically have S/N varying from 50:1 to 100:1. The analyses performed here use coadded spectra obtained with the HARPS spectrograph which has a resolution of 115\,000 \citep{may03}, and produces spectra with high S/N, as shown in Table~\ref{S/N}. Thus these spectra are more suitable for determining stellar parameters. The spectra were reduced using the standard HARPS data reduction software and the observation details for these spectra are discussed in \citet{tr10}, \citet{qu10} and \citet{gi09}.

The paper is divided as follows: section 2 outlines the methods used in the analyses, section 3 gives the parameters of the standard stars that were used to test the methods, section 4 details a discussion of the results, and the conclusion is given in section 5.

\begin{table}
\centering
\caption{Details of the HARPS spectra used in the present work}
\begin{tabular}{l l l l} \hline
Star & V mag & No. of spectra & S/N\\ \hline
WASP-2 & 11.98 & 32 & 100  \\
WASP-4 & 12.5 & 32 & 100  \\
WASP-5 & 12.3 & 34 & 125  \\
WASP-6 & 11.9 & 44 & 125   \\
WASP-7 & 9.5 & 20 &  235 \\
WASP-8 & 9.79 & 80 & 430  \\
WASP-15 & 11.0 & 51 & 240\\
WASP-16 & 11.3 & 77 & 175\\
WASP-17 & 11.6 & 60 & 230 \\
WASP-18 & 9.3 & 21 & 270  \\
WASP-19 & 12.59 & 36 & 125\\
\hline
\label{S/N}
\end{tabular}
\end{table}

\section{Method}
The spectral synthesis package used was{\sevensize\ UCLSYN} (University College London SYNthesis; \citealt{sd88}; \citealt{sm92}; \citealt*{ssd01}). {\sevensize\ ATLAS}9 models atmospheres without convective overshooting are used \citep*{cgk97} and local thermodynamic equilibrium (LTE) is assumed. 

\subsection{Line list}
It is impractical to individually measure every line in a spectrum, and so an appropriate selection of lines must be chosen from a high S/N spectrum, such as the Kurucz solar atlas \citep{ku84} as was used here. Lines were also selected from the HARPS Procyon (HD 61421) spectrum, so that these could be used for hotter stars in addition to the main line list. The line list was constructed by prioritising unblended lines where possible, and these comprise around one third of the list. Some lines had blending components that could be resolved via spectral synthesis, and these were included to increase the number of lines. Unresolved blends were included when there was a paucity of lines for a particular element. In this case, a restriction was imposed to ensure that the blending component comprised no more than five per cent of the overall equivalent width (EW), otherwise the line was deemed to be too severely blended to be of use. Lines which were blended with the same element in a different ionisation state were rejected, even if the weaker component was less than five per cent, as lines blended with other ionisation states could bias our estimate of the \teff. Unresolved blends were rejected for Fe, as there are still a sufficient number of Fe lines remaining after these blends are excluded. It is important to include as many Fe~{\sc ii} lines as possible, as these are essential when determining \logg\ via the ionisation balance method. 

A multitude of lines will not necessarily decrease the abundance error, especially if the additional lines are of poor quality. According to \citet{ku02}, abundance errors are likely to increase if there is a wide range of line strengths in a line list. An accurate abundance can theoretically be determined from a single weak line, as weak lines (less than 30 m{\AA}) are less affected by damping parameters and microturbulence (\mictrb). As such, including strong lines can increase the abundance errors. However, it is not possible to use only weak lines as there are not a sufficient number of them, and thus strong lines were also included in the list. The EWs for the Fe~{\sc i} lines in the HARPS solar spectrum range from 5.6 to 133.1 m{\AA} (although the strongest lines were often rejected at a later stage) and the lower level excitation potential ($\chi$) ranges from 0.052 to 5.033 eV. 

Atomic data was mainly taken from the VALD database \citep{ku99}, however damping parameters were obtained from \citet{kb95} when no data was available in VALD. In addition, the default oscillator strength (\loggf) value was not always used for certain elements (e.g. Cr~{\sc ii}, Ti~{\sc ii}), as the value given resulted in abundances that were highly inconsistent with the assumed solar abundances. For instance, using the default VALD values yields a $\log A$(Cr~{\sc ii}) of 5.80, which is 0.16 dex higher than the \citet{as09} abundance. However, using alternative \loggf\ values from VALD results in a $\log A$(Cr~{\sc ii}) of 5.66. In these cases the different source of \loggf\ from within VALD was used, so that a more suitable solar abundance could be obtained. The \loggf\ values were also supplemented from other sources where possible (\citealt{fw06}; \citealt{mb09}). The choice of \loggf\ can greatly influence the abundance obtained, resulting in a wide range of abundances obtained for a particular element, even in the Sun. For example, the solar Fe abundance varies between 7.41 and 7.56 throughout the literature \citep{ma11}, and the solar Mn ranges between 5.23 and 5.46 \citep{bg07}.

All lines were cross referenced with the NIST database to check the reliability of \loggf\ values. Any lines with an ``E'' rating, i.e. with an uncertainty of greater than $\pm$~50 per cent \citep{fw06}, were rejected. However, it should be noted that not all lines in the list were present in the NIST database, so some lines that were included in our line list could still have large uncertainties in the \loggf\ values. The line list is given in Table~\ref{lines}, showing the NIST rating where available.

\begin{table}
\begin{minipage}{80mm}
\caption{A sample of the line list used in the analyses of the HARPS spectra.}
\begin{tabular}{ccccc} \hline 
Element & Wavelength & $\chi$ & \loggf\ & NIST rating \\ \hline
Co~{\sc ii} & 4516.633 & 3.459 & -2.562 & \\
Fe~{\sc ii} & 4546.467 & 4.186 & -2.510 &  \\
Ca~{\sc i} & 4578.551 & 2.521 & -0.697 &  \\
Ti~{\sc ii} & 4583.409 & 1.165 & -2.870 & D \\
Cr~{\sc ii} & 4588.199 & 4.071 & -0.627 & D \\
Cr~{\sc ii} & 4592.049 & 4.074 & -1.221 & D \\
Fe~{\sc ii} & 4620.521 & 2.828 & -3.210 & C \\
Fe~{\sc i} & 4631.486 & 4.549 & -1.594 &  \\
Fe~{\sc ii} & 4656.981 & 2.891 & -3.600 & C \\
Fe~{\sc ii} & 4720.149 & 3.197 & -4.480 & D  \\
Mn~{\sc i} & 4739.087 & 2.941 & -0.490 & B  \\
\hline
\end{tabular}
The NIST rating is given for each line where available. The complete table is available as supplementary content in the online version of this paper.
\label{lines}
\end{minipage}
\end{table}

\subsection{Balmer lines as temperature indicators}
The Balmer lines can be used as an indication of \teff\ for stars cooler than 8000 K \citep{gr08}. The core to wing transition is the part of the line which is most sensitive to temperature changes \citep{ma10} and the outer wing broadening of \halpha\ is barely affected by other stellar parameters \citep{st07}. However, Balmer lines are not useful for determining temperature in the HARPS spectra, as the overlapping echelle orders are merged during pipeline reduction \citep{br10a}. This affects \halpha\ more than \hbeta, and makes normalisation extremely difficult. As such, the Balmer lines were only used to gain an initial estimate for \teff.

\subsection{Surface gravity from pressure-sensitive lines}
\label{Pressure}
Strong, pressure-broadened lines in the spectra were fitted in order to obtain a value for \logg. The Na~{\sc i} D lines were the main lines used, however it can be difficult to acquire the \logg\ from these lines if interstellar Na is present. The abundance used can also influence the \logg\ measurements. For example, there are only two suitable unblended Na lines in the spectra, other than the Na~{\sc i} D lines, which can be used for abundance determination. The lack of lines could give an erroneous abundance, which is then incorporated into the \logg\ determination. There are more Ca lines which can be used to determine abundance, which led to the Ca line at 6439 {\AA} also being used as a gravity diagnostic. However this line is not as useful in hotter stars, as the sensitivity to changes in \logg\ is reduced. The Mg~{\sc i} b lines were not used, as the only line that gives a reasonable value for \logg\ is the 5167 {\AA} line, and this is severely blended. 

\subsection{Line Broadening}
\label{Broadening}
Microturbulence and macroturbulence (\mactrb) are ad hoc broadening parameters that are needed in order fit the line profiles to the observed spectra. It was found that the abundance derived from strong lines differed from those calculated using weak lines, and thus introducing \mictrb\ would bring the abundances into agreement \citep{se34}. Microturbulence represents a small scale Gaussian velocity field, that is in added to the thermal Doppler broadening. It results in strong lines being spread over a larger wavelength interval and becoming saturated at larger EWs than if no \mictrb\ was included \citep{la06}. Macroturbulence only broadens the lines and does not change their line strengths. The size of the microturbulent cell is defined as being less than the mean free path of the photon, whereas \mactrb\ represents velocities that occur where the cell is larger than the unit optical depth \citep{ho03}. This division between micro and macroturbulence is crude, as in reality the motion would occur over a range of scales (\citealt{mu11}; \citealt{as00}; \citealt*{no09}). It has been shown that 3D models have no need for \mictrb\ and \mactrb, as convective energy transport can be included in the models \citep{as05}.

Microturbulence was determined from reducing the trend in EW with abundance. Weak lines should not be affected by \mictrb, therefore requiring that the abundances of strong lines equal those of the weak lines should result in a value for \mictrb. While the \citet{mag84} method of determining \mictrb\ using calculated EWs agrees with the observed EWs (within the errors) for spectra with S/N greater than 70 \citep{mu11}, there is still a small difference between the two methods for some stars. As such, the \citet{mag84} method was preferred for calculating the \mictrb. The scatter in abundances makes it difficult to get get a precision in \mictrb\ any better than 0.08 \kms. Also, obtaining the appropriate \mictrb\ value depends strongly on knowing the correct value of \teff\ and \logg. If one of these two parameters is incorrect then the \mictrb\ will be skewed. For instance, a change in \teff\ or \logg\ by 100 K or 0.1 dex respectively will result in a change in \mictrb\ of around 0.1 \kms.

Macroturbulence was calculated based on the \citet{br10a} calibration for stars with \teff\ below 6500 K, and this was extrapolated for stars greater than 6500 K. Rotational velocity (\vsini) was then fit for a selection of unblended Fe~{\sc i} lines. The radiative damping constant, Van der Waals damping constant, and the Stark broadening factor were input via the line list. The instrumental broadening, determined from the telluric lines at around 6300 \AA, was also included.

\subsection{Parameters from Fe lines}
The \teff, \logg\ and \mictrb\ of a star can be determined using Fe lines (\citealt{gv98}; \citealt*{br08}; \citealt*{sim00}; \citealt{so06}). Theoretically, this can be done with any element, but only FGK stars have a suitable number of Fe lines present to perform a precise analysis.

Fe~{\sc i} abundances will increase with increasing \teff. This temperature sensitivity is greater for low $\chi$ lines, and is almost negligible for high $\chi$ lines. Thus requiring that there is no trend between $\chi$ and abundance should yield the \teff\ of the star. The same principle can be applied to Fe~{\sc ii} lines, where in this case it is the high $\chi$ lines that are sensitive to \teff\ changes, however there are not usually a sufficient number of Fe~{\sc ii} lines present. The Fe~{\sc ii} abundance will increase with increasing \logg, where as the Fe~{\sc i} abundance is insensitive to \logg\ variations. Thus, requiring that the Fe~{\sc i} and Fe~{\sc ii} abundances agree should result in a value for \logg. These methods for determining the \teff\ and \logg\ from Fe lines are known as excitation and ionisation balance respectively \citep*{tos02}. 

The mean of the \logg\ determined from the ionisation balance and from the pressure-broadened lines was used as the overall value. The \logg\ from the ionisation balance agreed with the \logg\ from the pressure-broadened lines to within 0.3 dex. This difference is most likely due to the errors in abundance, discussed in Sect.~\ref{Pressure}, or due to \teff\ errors influencing the ionisation balance. \citet{br10b} also recommend the use of the Ca~{\sc i} 6122 and Ca~{\sc i} 6162 {\AA} lines, however we found that these lines often gave a \logg\ value that differed greatly from the \logg\ determined from the other pressure-broadened lines and the ionisation balance, therefore these lines were not included in the overall \logg\ estimation. 

Any obvious outliers in the plot of Fe abundance against EW or $\chi$ were removed, with a requirement that the abundance of the lines should be within 0.25 dex of the average abundance. This restriction allowed discrepant lines to be removed while still retaining a sufficient amount of lines needed to determine the stellar parameters. In general it was found that the same lines had abundances that were too high or too low, indicating that these lines had incorrect atomic data.

Microturbulence was determined by requiring a null dependence between Fe abundance and EW (see Sect.~\ref{Broadening}).

\subsection{Line measurements and continuum determination}
We calculated the abundances by using {\sevensize\ UCLSYN} to interactively measure the EWs of as many absorption lines as possible for each element. A direct numerical integration of the line profile was performed in order to determine the EW. A synthetic profile was then generated and compared to the observed profile shape. If the synthetic profile failed to produce an acceptable fit to the observed spectrum, for example the observed spectral line could have been broader than the synthetic spectral line, then a least squares method fit was used to obtain the EW.

In order to directly measure the EWs, a point on each side of the line is selected where the wing of the line reaches the continuum. However determining these points can be difficult, particularly for strong lines, and often results in an underestimate of the EW and thus the abundance. 

\begin{figure}
\centering
\includegraphics[height=\columnwidth,angle=270]{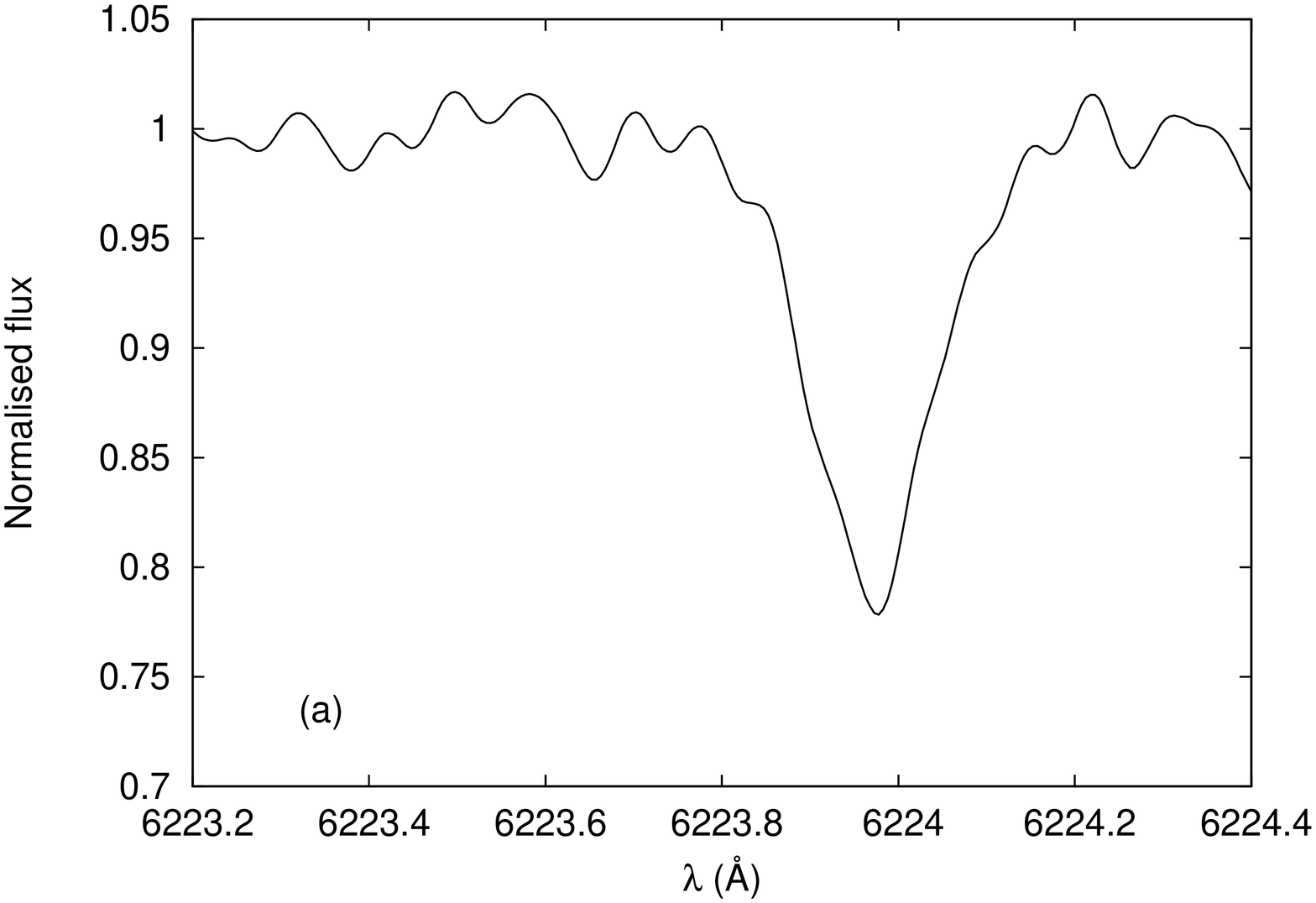}
\includegraphics[height=\columnwidth,angle=270]{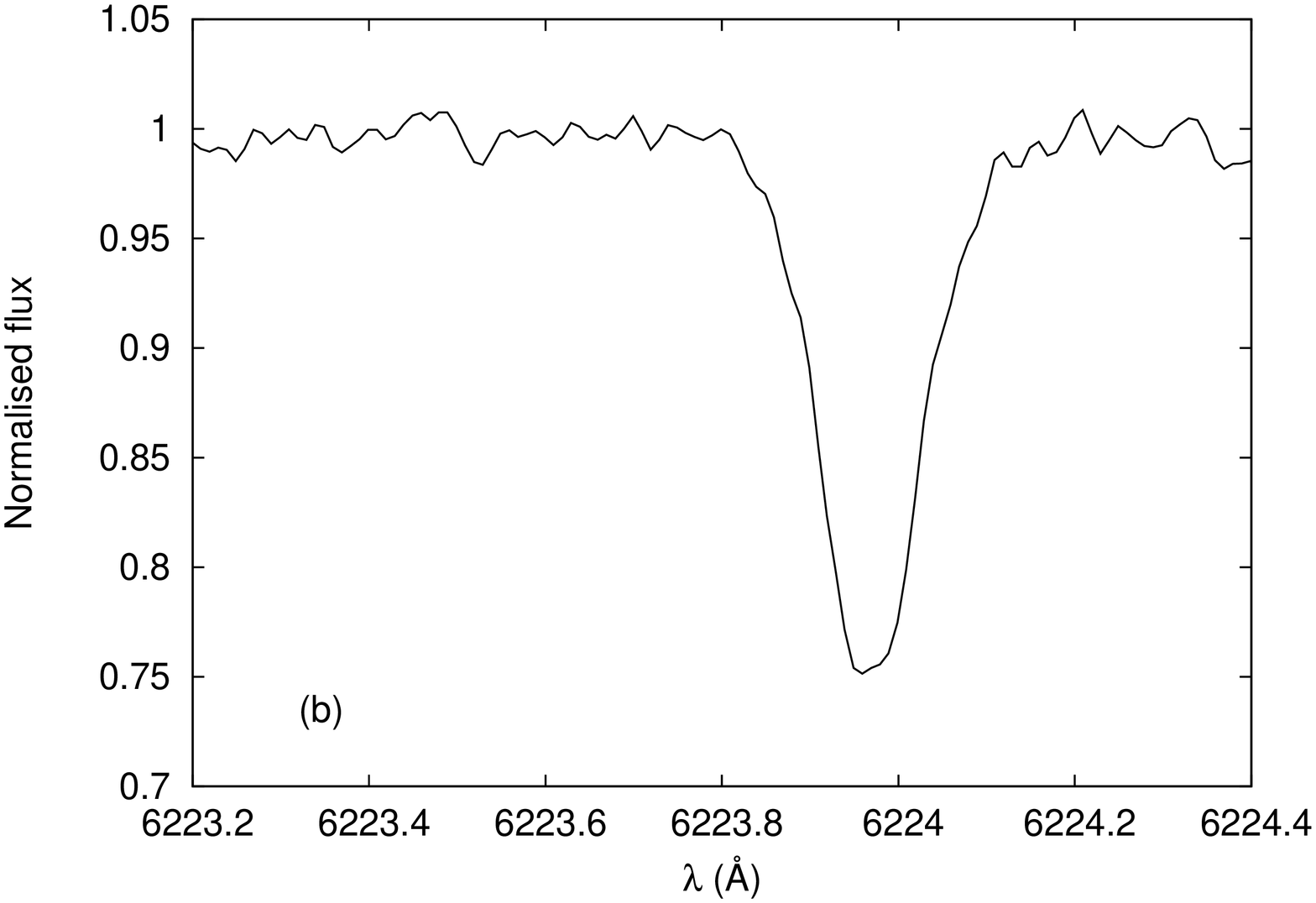}
\caption{A Ni~{\sc i} line in a CORALIE spectrum of WASP-16 with S/N of 70 is shown in the top panel. The noise in the continuum makes it difficult to normalise the spectrum, as well as creating problems with fitting the wings of the line. The same line is shown in the bottom panel with the HARPS spectrum (S/N 175) for comparison.}
\label{norm-cor}
\end{figure}

Uncertainties in continuum placement can also affect the measurement of EWs, and thus influence the abundance. The continuum was carefully normalised by eye over a small wavelength range for each line in order to ensure maximum precision. It was found that in spectra with low S/N (such as the CORALIE spectra), the noise makes the continuum placement difficult. In addition, the wings of the lines become lost in the noise, leading to an underestimate in the line strength. An example is shown in Fig.~\ref{norm-cor} for a line in the CORALIE WASP-16 spectrum (S/N of 70), as well as for the same line in the HARPS spectrum which has S/N of 175

\begin{figure}
\centering
\includegraphics[height=\columnwidth,angle=270]{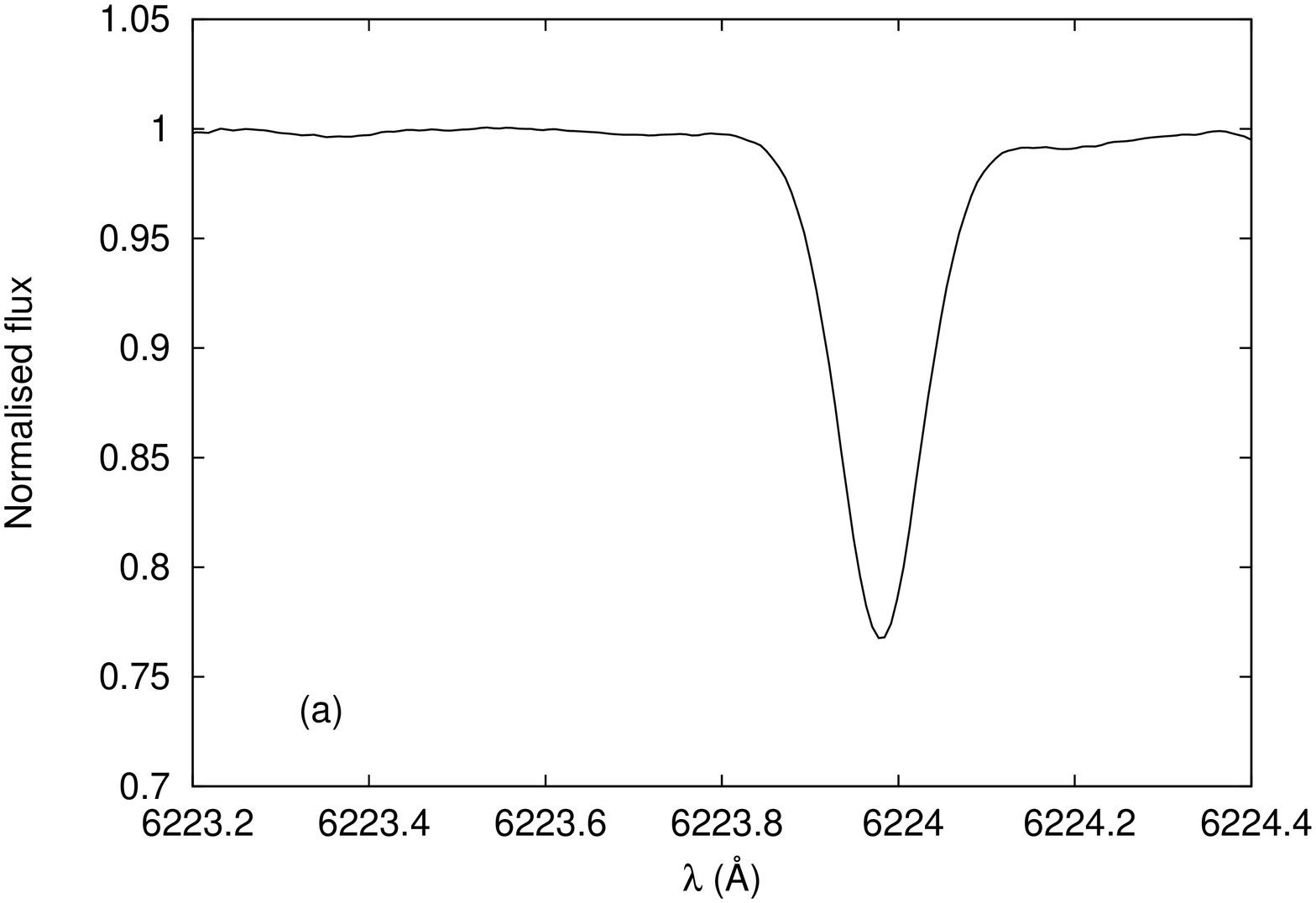}
\includegraphics[height=\columnwidth,angle=270]{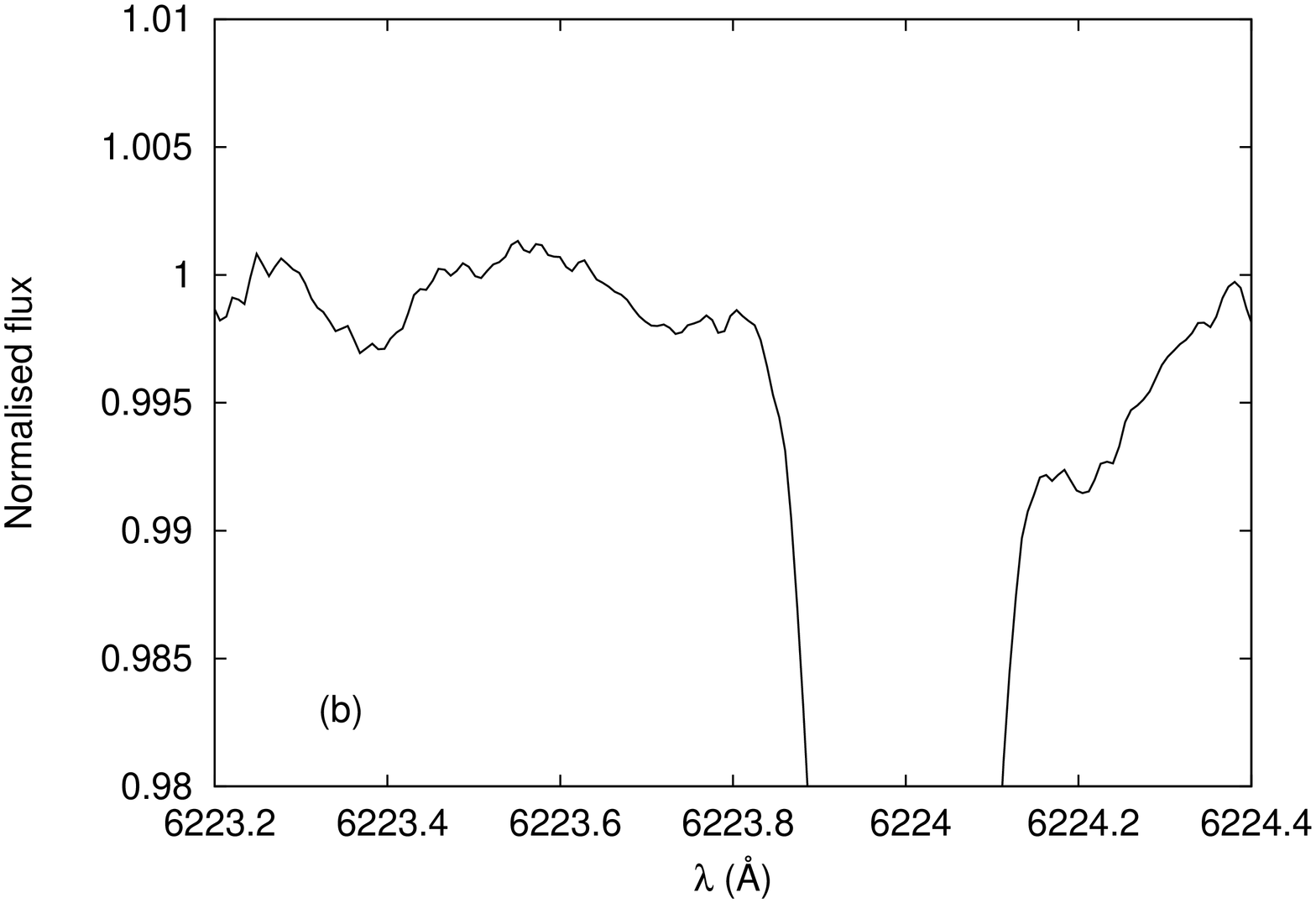}
\caption{The top panel shows the same Ni~{\sc i} line as Fig.~\ref{norm-cor}, but in the Kurucz solar atlas (S/N of 3000). The lower panel is a close up of the continuum showing that weak lines can cause issues even in high S/N spectra.}
\label{norm}
\end{figure}

A high S/N spectrum doesn't necessarily eliminate all problems associated with determining where the line wings meet the continuum. Extremely weak lines that are ordinarily lost in the noise at the continuum become evident, but these are often unidentified lines which makes them difficult to synthesise, as is shown in Fig.~\ref{norm} for the Kurucz solar atlas.

\section{Standard stars}
Before applying the method outlined above to the HARPS spectra of the WASP stars, it was important to verify that the method achieved acceptable results on well known stars such as the Sun and Procyon. Procyon was chosen as a hotter comparison star than the Sun, as many WASP stars have temperatures higher than the solar value. The parameters were derived from the Kurucz solar atlas (S/N $\sim$3000) and the HARPS sky spectrum (S/N $\sim$1000) for the Sun, and from a HARPS spectrum of Procyon. These parameters, along with literature values, are displayed in Table~\ref{stars}. The parameters of the Sun are well known, however the parameters of Procyon are less accurate. The mass, and thus the \logg, can be determined with a good deal of accuracy due to the binary nature of Procyon. However, there is still some disagreement as to the \teff, mainly due to different values of bolometric flux and angular diameter. The \teff\ values vary from 6516 $\pm$ 87 K \citep*{alk05}, 6530 $\pm$ 49 K \citep{al02} and 6591 $\pm$ 43 K \citep{ch12}. \citet{ca10} also obtained a value of 6626 $\pm$ 80 K using the Infrared Flux Method (IRFM; \citealt{bs77}; \citealt*{bps80}). An averaged \teff\ value is adopted for Table~\ref{stars}.

In this analysis the solar $\log A$(Fe) was determined to be 7.52 $\pm$ 0.08 (for the HARPS solar spectrum). This is in good agreement with the values of 7.50 $\pm$ 0.04 and 7.52 $\pm$ 0.06 found by \citet{as09} and \citet{ca11} respectively, but higher than the value of 7.45 $\pm$ 0.02 determined by \citet{mb09}.

\begin{table*}
\centering
 \begin{minipage}{150mm}
\caption{Parameters obtained for the Sun and Procyon.}
\begin{tabular}{l l l l l l } \hline
Parameter & Sun (Kurucz) & Sun (HARPS) & Sun (literature) & Procyon (HARPS) & Procyon (literature)\\ \hline
\teff (K) & 5760 $\pm$ 50 & 5775 $\pm$ 45 & 5777$^{a}$ & 6660 $\pm$ 95 & 6566 $\pm$ 65$^{b}$ \\
\logg	  & 4.42 $\pm$ 0.02 & 4.43 $\pm$ 0.02 & 4.44$^{a}$ & 4.05 $\pm$ 0.06 & 4.01 $\pm$ 0.03$^{c}$\\
$\log A$(Fe)  & 7.49 $\pm$ 0.06 & 7.52 $\pm$ 0.08 & 7.50 $\pm$ 0.04$^{d}$ & 7.48 $\pm$ 0.09 & 7.36 $\pm$ 0.03$^{e}$\\
$\mictrb$ (\kms) & 0.85 $\pm$ 0.08 & 0.75 $\pm$ 0.15 & 0.85$^{f}$ &  1.70 $\pm$ 0.08 & 2.2$^{e}$\\
\hline
\end{tabular}
$^{a}$ \citet{gr08}, $^{b}$ See text, $^{c}$ \citet{ch12}, $^{d}$ \citet{as09}, $^{e}$ \citet{al02}, $^{f}$~\citet{mag84}
\label{stars}
\end{minipage}
\end{table*}

\section{Results and discussion}

\begin{figure}
\centering
\includegraphics[height=\columnwidth,angle=270]{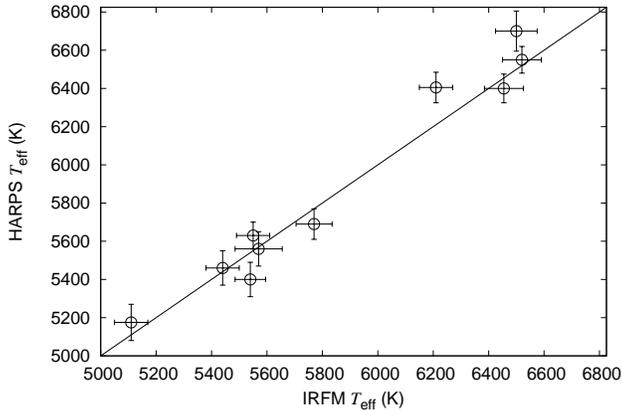}
\caption{Comparison of \teff\ from the HARPS spectra with the IRFM. The solid line depicts the 1:1 relationship.}
\label{teff-irfm}
\end{figure}

The updated \teff, \logg\ and $\log A$(Fe) values for the WASP stars are given in Table~\ref{teff-logg}, along with the initial spectroscopic \teff, \logg\ and $\log A$(Fe) from the discovery papers. These parameters were obtained mainly from the CORALIE spectra, except for WASP-2 which used a SOPHIE spectrum and WASP-8 which also used a HARPS spectrum. The IRFM \teff\ from \citet*{mks11} is also included, and the \teff\ derived from the HARPS spectra is compared with the IRFM \teff\ in Fig.~\ref{teff-irfm}. A 1:1 relationship shows that there is a good agreement between the two different methods, with the average \teff\ difference being 31 $\pm$ 110 K. However, it is possible that the spectroscopic temperatures are higher than the IRFM values, particularly as the \teff\ we determined for Procyon is higher than the average literature value. In the case of WASP-17, strong interstellar reddening is present which could affect the IRFM \teff.

Table~\ref{params} gives the \mictrb, \mactrb\ and \vsini. The mass and radius for each star are obtained from the \teff, \logg, and [Fe/H], based on the \citet{tag10} calibration. However, it should be noted that these masses and radii are only given as an example, and that stellar evolutionary models may give different results based on the same spectroscopic parameters.

The derived abundances are given in Table~\ref{abds} for all elements that have three or more usable spectral lines. The abundances obtained from the HARPS solar spectrum are largely consistent with the \citet{as09} solar values. However, there is a discrepancy with Co, and the error on Mn is large compared to the other elements. This may be due to hyperfine splitting as discussed in Section~\ref{errors}. Overall, it was found that the $\log A$(Fe) values derived from the HARPS spectra were an average of 0.09 $\pm$ 0.05 dex higher than the previous analyses. This is excluding WASP-2, WASP-4, and WASP-5, as there are large uncertainties in the initial values. In addition, the initial analysis of WASP-8 was also of a HARPS spectrum, so that the abundances agree between both examinations of the spectrum. Higher abundances are to be expected from the high S/N HARPS spectra, as the line profile wings are less likely to be underestimated due to noise on the continuum, however a detailed comparison is still difficult until more HARPS spectra have been analysed.

\begin{table*}
\centering
\scriptsize
\begin{minipage}{160mm}
\caption{Results from HARPS spectra compared with the initial analyses and the IRFM \teff.}
\begin{tabular}{l l l l l l l l} \hline
Star & HARPS \teff\ (K) & Initial \teff\ (K) & IRFM \teff\ (K) & HARPS \logg  & Initial \logg\ & HARPS $\log A$(Fe) & Initial $\log A$(Fe) \\ \hline
WASP-2 & 5175 $\pm$ 95 & 5200 $\pm$ 200$^{a}$ & 5110 $\pm$ 60 & 4.46 $\pm$ 0.12 & 4.3 $\pm$ 0.3$^{a}$   & 7.46 $\pm$ 0.10 &  $\sim$7.54$^{a}$ \\
WASP-4 & 5400 $\pm$ 90 & 5500 $\pm$ 150$^{b}$ & 5540 $\pm$ 55 & 4.47 $\pm$ 0.11 & 4.3 $\pm$ 0.2$^{b}$   & 7.42 $\pm$ 0.13 & 7.54 $\pm$ 0.20$^{b}$\\
WASP-5 & 5690 $\pm$ 80 & 5700 $\pm$ 150$^{c}$ & 5770 $\pm$ 65 & 4.28 $\pm$ 0.09 & 4.3 $\pm$ 0.2$^{c}$   & 7.63 $\pm$ 0.10 & 7.54 $\pm$ 0.20$^{c}$\\
WASP-6 & 5375 $\pm$ 65 & 5450 $\pm$ 100$^{d}$ & & 4.61 $\pm$ 0.07 & 4.6 $\pm$ 0.2$^{d}$	                & 7.35 $\pm$ 0.09 & 7.34 $\pm$ 0.10$^{d}$\\
WASP-7 & 6550 $\pm$ 70 & 6400 $\pm$ 100$^{e}$ & 6520 $\pm$ 70 & 4.32 $\pm$ 0.06 & 4.3 $\pm$ 0.2 $^{e}$  & 7.68 $\pm$ 0.06 & 7.54 $\pm$ 0.10$^{e}$\\
WASP-8 & 5560 $\pm$ 90 & 5600 $\pm$ 80$^{f}$ & 5570 $\pm$ 85 & 4.40 $\pm$ 0.09 & 4.5 $\pm$ 0.1$^{f}$    & 7.70 $\pm$ 0.11 & 7.71 $\pm$ 0.07$^{f}$\\
WASP-15 & 6405 $\pm$ 80 & 6300 $\pm$ 100$^{g}$ & 6210 $\pm$ 60 & 4.40 $\pm$ 0.11 & 4.35 $\pm$ 0.15$^{g}$ & 7.52 $\pm$ 0.10 & 7.37 $\pm$ 0.11$^{g}$\\
WASP-16 & 5630 $\pm$ 70 & 5700 $\pm$ 150$^{h}$ & 5550 $\pm$ 60 & 4.21 $\pm$ 0.11 & 4.5 $\pm$ 0.2$^{h}$   & 7.59 $\pm$ 0.10 & 7.55 $\pm$ 0.10$^{h}$\\
WASP-17 & 6700 $\pm$ 105 & 6550 $\pm$ 100$^{i}$ & 6500 $\pm$ 75 & 4.34 $\pm$ 0.23 & 4.2 $\pm$ 0.2$^{i}$  & 7.40 $\pm$ 0.10 & 7.29 $\pm$ 0.09$^{i}$\\
WASP-18 & 6400 $\pm$ 75 & 6400 $\pm$ 100$^{j}$ & 6455 $\pm$ 70 & 4.32 $\pm$ 0.09 & 4.4 $\pm$ 0.15$^{j}$  & 7.60 $\pm$ 0.08 & 7.54 $\pm$ 0.09$^{j}$\\
WASP-19 & 5460 $\pm$ 90 & 5500 $\pm$ 100$^{k}$ & 5440 $\pm$ 60 & 4.37 $\pm$ 0.14 & 4.5 $\pm$ 0.2$^{k}$   & 7.66 $\pm$ 0.11 & 7.56 $\pm$ 0.09$^{k}$\\
\hline
\end{tabular}
$^{a}$ \citet{acc07}, $^{b}$ \citet{wi08}, $^{c}$ \citet{an08}, $^{d}$ \citet{gi09}, $^{e}$ \citet{he09}, $^{f}$ \citet{qu10}, $^{g}$ \citet{we09}, $^{h}$ \citet{li09}, $^{i}$~\citet{an10}, $^{j}$ \citet{he09b}, $^{k}$ \citet{h10}
\label{teff-logg}
\end{minipage}
\end{table*}

\begin{table*}
\centering
\begin{minipage}{120mm}
\caption{Stellar parameters for the WASP stars using the HARPS spectra.}
\begin{tabular}{l l l l l l} \hline
Star & \mictrb\ (\kms) & \mactrb\ (\kms) & \vsini\ (\kms) & Mass ($M_{\sun}$) & Radius ($R_{\sun}$)  \\ \hline
WASP-2 & 0.70 $\pm$ 0.15 & 0.9 $\pm$ 0.3 & 1.9 $\pm$ 0.7 & 0.87 $\pm$ 0.07 & 0.90 $\pm$ 0.14   \\
WASP-4 & 0.85 $\pm$ 0.10 & 1.4 $\pm$ 0.3 & 3.4 $\pm$ 0.3 & 0.92 $\pm$ 0.07 & 0.92 $\pm$ 0.13   \\
WASP-5 & 0.75 $\pm$ 0.10 & 2.2 $\pm$ 0.3 & 3.9 $\pm$ 0.2 & 1.10 $\pm$ 0.08 & 1.24 $\pm$ 0.15   \\
WASP-6 & 0.70 $\pm$ 0.20 & 1.4 $\pm$ 0.3 & 2.4 $\pm$ 0.5 & 0.87 $\pm$ 0.06 & 0.77 $\pm$ 0.07   \\
WASP-7 & 1.40 $\pm$ 0.08 & 5.2 $\pm$ 0.3 & 18.1 $\pm$ 0.2 & 1.34 $\pm$ 0.09 & 1.32 $\pm$ 0.11  \\
WASP-8 & 0.95 $\pm$ 0.15 & 1.9 $\pm$ 0.3 & 2.7 $\pm$ 0.5 & 1.04 $\pm$ 0.08 & 1.05 $\pm$ 0.12    \\
WASP-15 & 1.15 $\pm$ 0.08 & 4.6 $\pm$ 0.3 & 4.9 $\pm$ 0.4 & 1.23 $\pm$ 0.09 & 1.15 $\pm$ 0.16   \\
WASP-16 & 0.85 $\pm$ 0.10 & 2.1 $\pm$ 0.3 & 2.5 $\pm$ 0.4 & 1.09 $\pm$ 0.09 & 1.34 $\pm$ 0.20   \\
WASP-17 & 1.40 $\pm$ 0.10 & 5.8 $\pm$ 0.3 & 9.8 $\pm$ 1.1 & 1.29 $\pm$ 0.12 & 1.27 $\pm$ 0.38 \\
WASP-18 & 1.15 $\pm$ 0.08 & 4.6 $\pm$ 0.3 & 10.9 $\pm$ 0.7 & 1.28 $\pm$ 0.09 & 1.29 $\pm$ 0.16 \\
WASP-19 & 1.00 $\pm$ 0.15 & 1.6 $\pm$ 0.3 & 5.1 $\pm$ 0.3 & 1.01 $\pm$ 0.08 & 1.07 $\pm$ 0.19  \\
\hline
\end{tabular}
The stellar mass and radius are estimated using the \citet{tag10} calibration, and \mactrb\ is calculated from the \citet{br10a} calibration.
\label{params}
\end{minipage}
\end{table*}

\begin{table*}
 \begin{minipage}{220mm}
\scriptsize
\caption{Abundances for the HARPS solar spectrum and the WASP stars.}
\begin{tabular}{ccccccccccc} \hline
Star & $\log A$(Ca) & $\log A$(Sc) & $\log A$(Ti) & $\log A$(V) & $\log A$(Cr) & $\log A$(Mn) & $\log A$(Co) & $\log A$(Ni) & $\log A$(Y) \\ \hline
Sun &   6.33 $\pm$ 0.09 & 3.10 $\pm$ 0.07 & 4.95 $\pm$ 0.08 & 3.92 $\pm$ 0.06 & 5.65 $\pm$ 0.06 & 5.50 $\pm$ 0.14 & 4.84 $\pm$ 0.11 & 6.24 $\pm$ 0.06 & 2.24 $\pm$ 0.07\\
Procyon & 6.37 $\pm$ 0.09 & 3.12 $\pm$ 0.12 & 4.98 $\pm$ 0.10 & 3.93 $\pm$ 0.03 & 5.65 $\pm$ 0.11 & 5.23 $\pm$ 0.06 & 4.83 $\pm$ 0.10 & 6.20 $\pm$ 0.08 & 2.28 $\pm$ 0.07 \\ 
WASP-2  & 6.36 $\pm$ 0.08 & 3.15 $\pm$ 0.07 & 5.00 $\pm$ 0.08 & 4.07 $\pm$ 0.08 & 5.64 $\pm$ 0.09 & 5.53 $\pm$ 0.11 & 5.08 $\pm$ 0.20 & 6.18 $\pm$ 0.09\\
WASP-4  & 6.25 $\pm$ 0.12 & 3.09 $\pm$ 0.14 & 4.87 $\pm$ 0.09 & 3.86 $\pm$ 0.06 & 5.59 $\pm$ 0.09 & 5.53 $\pm$ 0.13 & 4.87 $\pm$ 0.06 & 6.14 $\pm$ 0.13 & 2.14 $\pm$ 0.09\\
WASP-5  & 6.49 $\pm$ 0.09 & 3.32 $\pm$ 0.16 & 5.07 $\pm$ 0.12 & 4.07 $\pm$ 0.08 & 5.77 $\pm$ 0.08 & 5.84 $\pm$ 0.09 & 5.11 $\pm$ 0.07 & 6.37 $\pm$ 0.10 & 2.30 $\pm$ 0.14\\
WASP-6  & 6.19 $\pm$ 0.12 & 3.02 $\pm$ 0.10 & 4.83 $\pm$ 0.08 & 3.80 $\pm$ 0.09 & 5.50 $\pm$ 0.10 & 5.27 $\pm$ 0.12 & 4.67 $\pm$ 0.14 & 6.04 $\pm$ 0.12 & \\
WASP-7  & 		  &	  	    & 5.26 $\pm$ 0.24 &		        & 5.77 $\pm$ 0.05 &		    &		      & 6.37 $\pm$ 0.10\\
WASP-8  & 6.56 $\pm$ 0.08 & 3.23 $\pm$ 0.06 & 5.14 $\pm$ 0.10 & 4.19 $\pm$ 0.09 & 5.84 $\pm$ 0.05 & 5.66 $\pm$ 0.06 & 5.21 $\pm$ 0.17 & 6.49 $\pm$ 0.11 & 2.37 $\pm$ 0.09\\
WASP-15 & 6.31 $\pm$ 0.09 & 3.17 $\pm$ 0.10 & 4.97 $\pm$ 0.07 & 	        & 5.62 $\pm$ 0.08 & 5.30 $\pm$ 0.08 & 4.77 $\pm$ 0.10 & 6.12 $\pm$ 0.07 & 2.29 $\pm$ 0.08 \\
WASP-16 & 6.44 $\pm$ 0.12 & 3.14 $\pm$ 0.13 & 5.03 $\pm$ 0.10 & 4.00 $\pm$ 0.11 & 5.73 $\pm$ 0.10 & 5.67 $\pm$ 0.16 & 5.10 $\pm$ 0.16 & 6.41 $\pm$ 0.11 & 2.12 $\pm$ 0.05\\
WASP-17 & 6.27 $\pm$ 0.19 & 		    & 4.91 $\pm$ 0.06 & 		& 5.53 $\pm$ 0.05 & 5.19 $\pm$ 0.09 & 		      & 6.06 $\pm$ 0.10 & 2.20 $\pm$ 0.11\\
WASP-18 & 6.48 $\pm$ 0.12 & 		    & 5.07 $\pm$ 0.06 & 3.94 $\pm$ 0.08 & 5.75 $\pm$ 0.10 & 5.51 $\pm$ 0.15 & 		      & 6.24 $\pm$ 0.06 & 2.41 $\pm$ 0.10\\
WASP-19 & 6.51 $\pm$ 0.13 & 3.29 $\pm$ 0.09 & 5.15 $\pm$ 0.11 & 4.16 $\pm$ 0.08 & 5.81 $\pm$ 0.08 & 5.78 $\pm$ 0.21 & 5.12 $\pm$ 0.10 & 6.40 $\pm$ 0.08 & 2.39 $\pm$ 0.16\\
\hline
\end{tabular}
\label{abds}
\end{minipage}
\end{table*}

In addition to using abundances determined from the EWs measured using{\sevensize\ UCLSYN}, a differential analysis with respect to the Sun and Procyon was also performed. Both a line by line  differential analysis was used, as well as an alternate method where the solar $\log A$(Fe) was fixed, and the \loggf\ values were adjusted accordingly. It was found that the{\sevensize\ UCLSYN} abundances agree with the differential abundances to within 0.04 dex, and that the scatter is reduced for the differential abundances. However, a differential analysis is best performed when the stars have almost identical parameters to that of the reference star \citep{ta05}, and adjusting the solar \loggf\ values assumes that an accurate solar $\log A$(Fe) is known. As such, the abundances obtained from the non-differential analyses were retained.

\subsection{Errors}
\label{errors}
Errors  in \teff\ were calculated from the 1-$\sigma$ variation in the slope of abundance against excitation potential, and range between 70 and 105 K. These values are consistent with \citet{twh08}, who suggest that \teff\ errors should not fall below 50K, despite the fact that some automated spectroscopic analyses often give errors that are much lower than this. To justify this decision they cite, for example, the difference of around 100 K between excitation equilibrium measurements and the IRFM determined by \citet{rm04}. \citet{mks11} also performed a comparison between spectroscopic methods and the IRFM, which supports that temperature errors should not be any lower than 50 K.

Variations in \teff, \logg\ and \mictrb\ can affect the elemental abundances. Table~\ref{sun-abderr} and Table~\ref{pro-abderr} list the uncertainties when the stellar parameters are varied by their average errors for the HARPS solar spectrum and Procyon respectively. Elements such as V are more sensitive to changes in temperature than others due to a restricted range in excitation potential. Therefore a large error on \teff\ will give V a larger error than the other elements. For certain elements, additional sources of uncertainty have to be considered. Abundances can be overestimated for Mn and Co, as the line profiles are altered due to hyperfine splitting \citep{sc11}.

A change in \teff\ will affect the ionisation balance, for example an increase in \teff\ of 100 K will result in \logg\ increasing by 0.1 dex \citep{br09}. This was accounted for by varying the \teff\ by 1-$\sigma$ when using the ionisation balance method. The \logg\ determined from the pressure-broadened lines is also dependent on the abundance, as an increased abundance will cause the line to be stronger. The abundances were thus varied by 1-$\sigma$ when determining \logg\ from fitting the Na~{\sc i} D and Ca~{\sc i} lines.

While there is some reduction in errors of stellar parameters from the original low S/N spectra, the errors are still high considering the S/N of the HARPS spectra. We calculated the average random errors in \teff, \logg, \mictrb\ and $\log A$(Fe) to be 83 K, 0.11 dex, 0.11 \kms\ and 0.10 dex respectively for this sample of stars. The scatter in Fe abundances due to uncertainties in atomic data can influence the $\log A$(Fe) as well as \teff, \logg\ and \mictrb.

Systematic errors in the EWs can be investigated by comparing the EWs of the same star between two different spectrographs \citep{g07}. Figure~\ref{errors} shows the EWs of the Sun measured from the Kurucz solar atlas plotted against the EWs from the HARPS solar spectrum, along with a 1:1 relationship. The EWs agree between both spectra for weak lines, however the HARPS EWs seem to be weaker than the Kurucz solar atlas for stronger lines. There is no significant offset evident when comparing the results between the HARPS solar spectrum and the Kurucz solar atlas, as seen in Table~\ref{stars}. In addition, the strongest lines are usually culled prior to the final analyses. Therefore any systematic errors present should have a negligble effect on the stellar parameters.

\begin{figure}
\centering
\includegraphics[height=\columnwidth,angle=270]{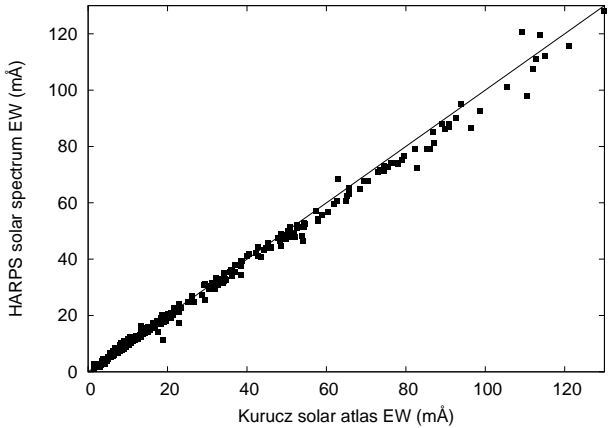}
\caption{Comparison of EWs measured in two different solar specta. The solid line depicts the 1:1 relationship.} 
\label{errors}
\end{figure}

\begin{figure}
\centering
\includegraphics[height=\columnwidth, angle=270]{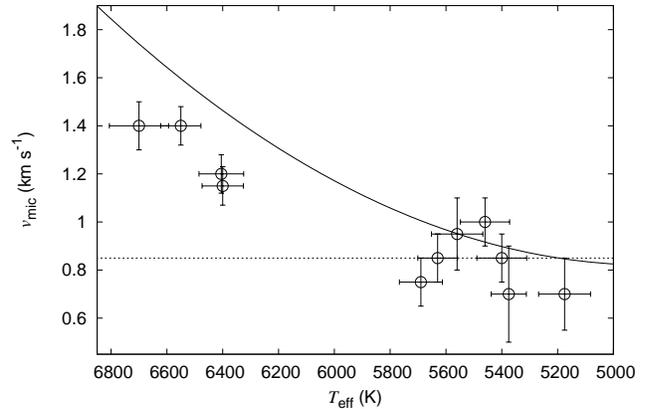}
\caption{Microturbulence is shown to increase with temperature. The Bruntt et al. (2010a) calibration is marked by the curved solid line, while the Valenti \& Fischer (2005) fixed value of 0.85 \kms\ is represented by the horizontal dashed line.}
\label{mict}
\end{figure}

\begin{table}
\centering
\caption{Abundance uncertainties for the HARPS solar spectrum.}
\begin{tabular}{l l r @{.} l r @{.} l } \hline
Element & $\Delta$ \teff\  & \multicolumn{2}{l}{$\Delta$~\logg} & \multicolumn{2}{l}{$\Delta$ \mictrb} \\ 
 & +83 K & +0&11 dex & +0&11 \kms\ \\ \hline
{[Fe/H]} & 0.02 & 0&00 & -0&02 \\ 
{[Ca/H]} & 0.05 & -0&03 & -0&02 \\
{[Sc/H]} & 0.01 & 0&04 & 0&00 \\
{[Ti/H]} & 0.05 & 0&01 & -0&01 \\
{[V/H]} & 0.09 & 0&00 & 0&00\\
{[Cr/H]} & 0.04 & 0&01 & -0&01\\
{[Mn/H]} & 0.07 & 0&00 & -0&02\\
{[Co/H]} & 0.06 & 0&01 & 0&00\\
{[Ni/H]} & 0.04 & 0&00 & -0&01\\
{[Y/H]} & 0.00 & 0&05 & -0&01\\
\hline
\end{tabular}
\label{sun-abderr}
\end{table}

\begin{table}
\centering
\caption{Abundance uncertainties for Procyon.}
\begin{tabular}{l l r @{.} l r @{.} l } \hline
Element & $\Delta$ \teff\  & \multicolumn{2}{l}{$\Delta$~\logg} & \multicolumn{2}{l}{$\Delta$ \mictrb} \\ 
 & +83 K & +0&11 dex & +0&11 \kms\ \\ \hline
{[Fe/H]} & 0.03 & 0&01 & -0&01 \\
{[Ca/H]} & 0.05 & -0&01 & -0&02 \\
{[Sc/H]} & 0.04 & 0&04 & 0&00 \\
{[Ti/H]} & 0.03 & 0&02 & -0&01 \\
{[V/H]} & 0.05 & 0&01 & 0&00 \\
{[Cr/H]} & 0.02 & 0&02 & -0&02 \\
{[Mn/H]} & 0.11 & 0&06 & 0&05 \\
{[Co/H]} & 0.05 & 0&00 & 0&00 \\
{[Ni/H]} & 0.05 & 0&00 & 0&00 \\
{[Y/H]} & 0.03 & 0&04 & -0&01 \\
\hline
\end{tabular}
\label{pro-abderr}
\end{table}

\subsection{Microturbulence variations with \teff}
The \mictrb\ was found to increase with \teff, in agreement with \citet{br10a}, \citet{la09}, and \citet{s04}. It has been suggested that \mictrb\ should be fixed to the solar value of 0.85 \kms\ \citep{vf05}. However, while this assumption is valid for solar-like stars, it is not appropriate for hotter stars, as is shown in Fig.~\ref{mict}, and this could skew other stellar parameters for hotter stars.
 
\section{Summary and conclusion}
In an effort to improve the precision of stellar parameters of eleven WASP stars from the initial analyses, we have obtained parameters using high S/N HARPS spectra. We created a new line list using the VALD database, but also supplementing atomic data from other sources when necessary. The line list was created by selecting lines from the Kurucz solar atlas, and also from the HARPS Procyon spectrum to create a subset of lines for hotter stars. The line list was cross referenced with the NIST database in order to reject any E-rated lines, i.e. lines with a \loggf\ uncertainty greater than $\pm$ 50 per cent, however not all lines in the list were also present in the database.

We determined elemental abundances via measuring EWs, and a least squares fit of spectral lines was performed when the synthetic profile failed to agree with the observed spectrum. The continuum was normalised by eye over a small wavelength range for each line, however even for high S/N spectra it was found that some uncertainty still exists in continuum placement. We determined the \teff, \logg, and \mictrb\ from Fe lines by requiring that there is no trend present when abundance is plotted against either EW or excitation potential. It was also essential that the abundances of Fe~{\sc i} and Fe~{\sc ii} agreed. The \logg\ was also determined from the Na~D and Ca~{\sc i} line at 6439 \AA. We tested these methods on two different solar spectra and a spectrum of Procyon before being applied to the WASP stars. The masses and radii of the stars, which are needed to obtain planetary parameters, were calculated based on the \citet{tag10} calibration. 

We found that the $\log A$(Fe) values determined from the HARPS spectra were an average of 0.09 $\pm$ 0.05 dex higher than those obtained from the lower S/N spectra used in the initial analysis of the WASP stars. This is most likely due to the higher S/N of the HARPS spectra allowing for the wings of the spectral lines to be measured with greater precision. We discussed the importance of clearly stating the solar $\log(A)$ values. Stellar metallicities are often given relative to the Sun, but the solar value used is not always stated, which can cause additional discrepancies when comparisons are made between different methods. The average uncertainty in \teff, \logg, \mictrb\ and $\log A$(Fe) is 83 K, 0.11 dex, 0.11 \kms\ and 0.10 dex respectively, showing that there is a limit to the accuracy that can be achieved even using high S/N spectra.

\section*{Acknowledgments}
A.P.D. acknowledges support from EPSAM at Keele University. This research has made use of NASA's Astrophysics Data System.

\label{lastpage}

\end{document}